\documentclass{pnastwo}

\url{www.pnas.org/cgi/doi/10.1073/pnas.0709640104}
\copyrightyear{2008}
\issuedate{Issue Date}
\volume{Volume}
\issuenumber{Issue Number}
\begin{document}

\title{Counterfactual Definiteness and Bell's Inequality}

\author{Karl Hess\affil{1}{Center for Advanced Study, University of Illinois, Urbana, Illinois},
Hans De Raedt\affil{2}{
Department of Applied Physics,
Zernike Institute for Advanced Materials,
University of Groningen, Nijenborgh 4, NL-9747 AG Groningen, The Netherlands
},
Kristel Michielsen\affil{3}{
Institute for Advanced Simulation, J\"ulich Supercomputing Centre,
Forschungszentrum J\"ulich, D-52425 J\"ulich,
RWTH Aachen University, D-52056 Aachen,
Germany
}
}

\contributor{Member submission to the Proceedings of the National Academy of Sciences
of the United States of America}

\maketitle

\begin{article}
\begin{abstract}
Counterfactual definiteness must be used as at least one of the postulates or axioms that are necessary to derive Bell-type
inequalities. It is considered by many to be a postulate that is not only commensurate with classical physics (as for example
Einstein's special relativity), but also separates and distinguishes classical physics from quantum mechanics. It is the purpose
of this paper to show that Bell's choice of mathematical functions and independent variables implicitly includes
counterfactual definiteness and reduces the generality of the physics of Bell-type theories so significantly that no meaningful
comparison of these theories with actual Einstein-Podolsky-Rosen experiments can be made.
\end{abstract}

\keywords{Bell Inequality | Foundations of Quantum Mechanics | Foundations of probability}

\dropcap{B}ell's theorem \cite{BELL01} has an unusual standing among mathematical-physical theorems. No other theorem has ever
been discussed with respect to so many ``loopholes", physical situations that make it possible to escape the mathematical
strictures of the theorem. It is shown that the reason for this fact is that Bell's theorem is based on the postulate of
counterfactual definiteness. The postulate of counterfactual definiteness to derive Bell-type inequalities is clearly asserted in the books of
Peres \cite{PERE95} and Leggett \cite{LEGG85b}.
%and was summarized by Gerard 't Hooft \cite{tHOO15} with the statement ``..if
%you are not addressing the question of counterfactual measurements, you are not addressing Bell's theorem."

Some of Einstein's reasoning regarding Einstein-Podolsky-Rosen (EPR)
experiments contains also counterfactual realism and Einstein's special relativity is
counterfactually definite in the mathematical sense presented below. This fact may have contributed to the opinion that
counterfactual realism is the major defining trait of ``classical" theories. It will be shown, however, that great care must be
exercised with respect to the choice of independent variables in the argument of the functions that are used to formulate a
counterfactually definite physical theory. It will also be shown that the particular choice of variables, that are used for the
derivation of Bell's inequality and Bell's theorem, imposes significant restrictions to the physical situations that can be
described by Bell's functions and excludes dynamic processes of classical physics, no matter whether deterministic or
stochastic. To show this fact, we first repeat the main features of Bell's functions that describe
Einstein-Podolsky-Rosen-Bohm (EPRB) experiments and then connect them to a precise definition of counterfactual definiteness.

\section{EPRB experiments and Bell's functions representing them}

EPRB experiments are performed at two space-like separated locations.
The two particles of an entangled pair emanating from a source are spatially separated
and propagate to the space-like separated locations.
The properties of these particles are
measured by instruments that are described by a ``setting" such as the
direction of a polarizer or magnet which is characterized by a unit vector of three dimensional space denoted by
$\mathbf{j} = \mathbf{a}, \mathbf{b}, \mathbf{c},\ldots$.
Measurements of this type have been performed by a number of researchers and have had a checkered history with respect to
the results.
These, at first, contradicted and then confirmed quantum theory~\cite{GILD08}.
There are still significant deviations from quantum theory in current experiments, which are, however, mostly ignored~\cite{RAED13a}.
We proceed here by just stipulating that indeed these experiments showed a violation of the, by now, famous Bell inequality
and describe in the following only Bell's postulates and assumptions, thereby focusing on the simplest case
involving only three settings and not four, as used in actual experiments, see also~\cite{RAED16a}.
Bell's postulates and assumptions are considered by many researchers to be entirely general and valid for all EPR like experiments
and Gedanken-experiments as long as they can be described by classical physics such as Einstein's relativity.

Bell's classical-physics model for the system of measurement equipment and entangled pairs of the EPRB experiments is
constructed as follows~\cite{BELL01}.
He assumed that all experimental results, all data, can be described by using functions $A$ that map
the independent measurement results onto $\pm 1$ or the segment $[-1, +1]$ of the real axis. The variables in the
argument of the function always include the settings $\bf j = a, b, c,...$ and another variable, or set of variables, that Bell
denoted by $\lambda$. Bell then proceeded to present a proof of his now celebrated inequality:
\begin{equation}
\langle A(\mathbf{a}, \lambda)A(\mathbf{b}, \lambda) +
A(\mathbf{a}, \lambda)A(\mathbf{c}, \lambda) - A(\mathbf{b}, \lambda)A(\mathbf{c}, \lambda)\rangle \leq  +1
,
\label{5july15n1}
\end{equation}
where $\langle \cdot \rangle$ indicates the average over many measurements.
The left and right factor of each term correspond to the data taken at the two
corresponding space like separated measurement stations. The events of measurements and corresponding data are linked to clock
times of two synchronized laboratory clocks. Therefore, the functions $A$ as well as the variables ${\bf j}$ and $ \lambda$
must for each of the products correspond to pairs of clock times $t_n, t_n'$ where $n$ is the measurement number.

Note that Bell's original paper assigned to $\lambda$ only properties of the entangled pair. It is now generally assumed~\cite{BRAN87} that $\lambda$ may stand for a set of arbitrary physical variables
including space and time coordinates or even Einstein's space-time $\bf st$. Therefore, $\lambda$ may also describe some properties of the measurement equipment (in addition to the magnet or polarizer orientation $\bf j$), such as dynamical effects arising from many-body interactions of the entangled pair with the constituent particles and fields of the measurement equipment. Bell agreed with this assumption in his later work \cite{BELL01}.

It is the purpose of this paper to show that the postulate of counterfactual definiteness in conjunction with the use of a
setting variable $\bf j$ does not permit the introduction of general space and time related variables that describe the said
many body dynamics. Therefore, Bell's assumptions are not general enough to describe classical theories of EPRB experiments that
include dynamic processes involving the measurement equipment.

\section{Counterfactual reasoning and EPRB experiments}

Peres~\cite{PERE95} gave the following definition of counterfactual realism, which roughly agrees with the definition of
Leggett \cite{LEGG85b}. Peres claims, as does Leggett, not to use traditional concepts of mathematics and physics to start
with, but only ``what could have possibly been the results of unperformed experiments" and bases his definition of
counterfactual realism on the following statement:

\begin{center}
\medskip
\framebox{
\parbox[t]{0.8\hsize}{%
%%%%%
It is possible to imagine hypothetical results for any unperformed test, and to do calculations where these unknown results
are treated as if they were numbers.
%%%%%
}}
\medskip
\end{center}

We agree that it is possible, as a purely intellectual activity, to imagine hypothetical results for any unperformed tests.
However, without significant additional assumptions, it is not possible ``to do calculations where these unknown results are
treated as if they were numbers".
Here we encounter the so often unrecognized gulf between sense impressions, even just imagined
ones, and conceptual frame-works such as the axiomatic system of numbers or the probability theory of Kolmogorov. Peres, Leggett
and a majority of quantum information theorists did not and do not recognize that giant gulf, that giant separation, between events of nature, recorded
as data, and the axiomatic edifices of human thought.

If one wishes to treat hypothetical ``results" of unperformed tests as if they were numbers, one must be sure that these
abstractions at least follow the axioms of numbers. There are several steps necessary to connect the ``events" of the physical
world to numbers. Boole derived ultimate alternatives and a Boolean algebra while Kolmogorov's axiomatic system introduces an
event algebra and probability space. It is true that mathematicians often describe experimental situations or ideas about them
by the Kolmogorov framework and just postulate that a probability space and $\sigma$-algebra exists. It is known, however, from
the work of Boole \cite{BO1862} and Vorob'ev \cite{VORO62} that a given particular set of variables may not be able to describe
certain correlations in any given set of data.

In more elementary terms, we have to consider the following facts. If we perform ``calculations where these unknown results are
treated as if they were numbers", then we must use the mathematical concept of functions or something equivalent in order to
link the imagined but possible tests with numbers. A one to one correspondence of the possible tests and the numbers needs to be established
and it needs to be shown that no logical-mathematical contradictions arise from such procedure. If no such correspondence
exists, then the ``purely intellectual activity" is nothing more than child's play and the mathematical abstractions of such
activity can certainly not be treated as if they were numbers with some relation to physics.

Take any set of data derived from measurements on spin-$1/2$ particles with Stern-Gerlach magnets, that lists the
measured spins as ``up" or ``down" together with magnet settings
${\mathbf j}={\mathbf a}, {\mathbf b}, {\mathbf c}, \ldots$.
Can we replace ``up" with $+1$ and ``down"
with $-1$ and expect that the so obtained set follows the axioms of integers?
The ``trespass" to deal with tests as if they were numbers
has been committed by several textbook authors, in particular by Peres~\cite{PERE95} and Leggett~\cite{LEGG85b}.
This point appears in clear relief, if we write down the data according to the way in which they are imagined to be taken
in testing e.g. the Bell-type inequality.
The data are recorded in pairs corresponding to detector-events that are registered together with equipment settings
and the clock times of synchronized laboratory clocks.
Thus we obtain data lists of the kind:
$%\begin{equation}
(D_{{\mathbf j}_1}^{t_1}, D_{{\mathbf j}_1^\prime}^{t_1^\prime}),
(D_{{\mathbf j}_2}^{t_2}, D_{{\mathbf j}_2^\prime}^{t_2^\prime}),\ldots,
(D_{{\mathbf j}_N}^{t_M}, D_{{\mathbf j}_{M}^\prime}^{t_{M}^\prime}),
%\label{5aug15n1}
$ %\end{equation}
the ${\mathbf j}_n, {\mathbf j}_n'$ representing the randomly chosen setting pair and $t_n, t'_n$ denoting the times of measurement.
The number of times that the setting $(\mathbf{a},\mathbf{b})$, $(\mathbf{a},\mathbf{c})$,
and $(\mathbf{b},\mathbf{c})$ was chosen is denoted by
$N_{\mathbf{a},\mathbf{b}}$, $N_{\mathbf{a},\mathbf{c}}$, and $N_{\mathbf{b},\mathbf{c}}$, respectively.
The total number of pairs is then $M=N_{\mathbf{a},\mathbf{b}}+N_{\mathbf{a},\mathbf{c}}+N_{\mathbf{b},\mathbf{c}}$.
One cannot do justice to the number of different data-pairs by using models with three pairs of mathematical symbols such as
$A_{\mathbf{a}}, A_{\mathbf{b}}$, $A_{\mathbf{a}}, A_{\mathbf{c}}$ ,
and $A_{\mathbf{b}}, A_{\mathbf{c}}$ as they are used in Bell-type proofs.
One runs into problems even if one regards these mathematical symbols as ``variables"
(such as Boolean variables~\cite{HESS15a}) and not just as numbers;
the reason being that one cannot cover all the different possible correlations of the data by such few variables.
If we admit the two values $+1$ and $-1$ for the variables at different times of the same experiment,
then we obtain $N_{\mathbf{a},\mathbf{b}}+1$ different values for the sum of the pair product
$\sum_{n=1}^{M} \delta_{{\bf j}_n,\mathbf{a}}\delta_{{\bf j}_n^\prime,\mathbf{b}} D_{\mathbf{a}}^{t_n}D_{\mathbf{b}}^{t^\prime_{n}}$.
If we have three such sums with all independent variables, the number of possibilities is
$(N_{\mathbf{a},\mathbf{b}}+1)(N_{\mathbf{a},\mathbf{c}}+1)(N_{\mathbf{b},\mathbf{c}}+1)\approx (M/3 + 1)^3$ for
$M$ sufficiently large.
In contrast, we have for the Bell type variables
$A_{\mathbf{a}}, A_{\mathbf{b}}$, $A_{\mathbf{a}}, A_{\mathbf{c}}$, and $A_{\mathbf{b}}, A_{\mathbf{c}}$
only about $(M/3 +1)^2$ independent choices of all possible different correlations of possible
outcomes of these variables. This fact arises from Bell's description of $3M$ different {\bf pairs} of measurements
($6M$ measurements) by only 3 different variables and
represents another typical trespass that is explicitly made in both the book of Peres~\cite{PERE95} and Leggett~\cite{LEGG85b}:
they use a model with a severe restriction of choices before any physics is introduced
and thus``overburden" their variables in a way which cannot do justice to the complexity of the data.
In real EPRB experiments, one uses four not three different randomly chosen settings~\cite{ASPE82b,WEIH98}
but the above argument equally holds for this case, with $(M/3 + 1)^3$ and $(M/3 + 1)^2$ being
replaced by $(M/4 + 1)^4$ and $(M/4 + 1)^3$ for $4M$ different pairs ($8M$ measurements), respectively.

This more subtle problem, a well known problem in the area of computer simulations,
reveals once more the enormous gulf between data and mathematical abstractions that describe the data.
In the framework of Boole~\cite{HESS15a},
we need to be sure that the data can be described by ultimate alternatives (the Boolean variables)
and in the framework of Kolmogorov we must be sure to deal with random variables (functions on a Kolmogorov probability space).
But how can we be sure?
As a minimum requirement we need to introduce functions, with sufficiently many physical variables in their arguments,
to enable the description of all the possible correlations
and to guarantee a one to one correspondence of mathematical abstractions and the massive amount of data.

To describe EPRB experiments in the general way that Bell intended and purported to actually have done, we need to
introduce functions $A$ with variables additional to $\bf j$ in their argument (or indexes, see below).
We need to have variables such as $t_n, s_n, {\bf st}_n,\ldots$ that are taken out of the realm of Einsteinian physics
and do indeed guarantee the one to one correspondence to the data.
For example, we may need to include $t_n$, the time of measurement at one location and $s_n$
representing any property of the objects emanating from the source.
It may also be necessary to include a more general four dimensional space-time vector ${\mathbf st}_n$
instead or in addition to the measurement time $t_n$ and we include it here just for completeness.
This way we obtain functions $A = A({\bf j}_n, t_n, s_n, {\bf st}_n,\ldots)$.
%\begin{equation}
%$A = A({\bf j}_n, t_n, s_n, {\bf st}_n,\ldots).
%\label{feb5n1}
%\end{equation}

Some may ask whether that is not precisely what Bell used by introducing his
$\lambda$ that, as he claimed \cite{BELL01}, can stand for any set of variables and, therefore, also for the set
$(t_n, s_n, {\bf st}_n,\ldots)$.
We thus may have
$A = A({\bf j}_n, t_n, s_n, {\bf st}_n,\ldots)= A({\bf j}_n, \lambda_n)$.
%:
%\begin{equation}
%A = A({\bf j}_n, t_n, s_n, {\bf st}_n,\ldots)= A({\bf j}_n, \lambda_n).
%\label{march1315n1}
%\end{equation}
Indeed it is true that this is what Bell claimed. However, as we will see below his claim is
incorrect, because he and followers have postulated complete independence of $\lambda$ and $\bf j$ and thus postulated counterfactual definiteness in conjunction with the setting variable $\bf
j$ according to the precise definition given in the next section. Einstein locality does not require independence of $\lambda$ of the local setting (see corresponding section).

Note that quantum mechanics does not use any setting-type of variable as independent variable in the argument of the
wave-function. There, the setting-type variables label the operators. A helpful discussion of
explicit and implicit assumptions of Bell, with emphasis of the mathematical structure and consistency, was given by
Khrennikov~\cite{KHRE09}.

\section{Mathematical definition of counterfactual definiteness and Bell's inequality}

Counterfactual definiteness requires the following. We must be able to describe a
measurement or test by using a given set of variables in the argument of the function $A$, and thus for example a setting
$\mathbf{j} = \mathbf{b}$.
Then, we must also be able to reason that we could have used instead of setting $\bf b$ the setting $\bf c$ and would have
obtained the outcome corresponding to the value of $A$, now calculated with setting $\bf c$ and all other variables in its
argument unchanged. Although this type of reasoning is not permitted in the courts of law, its mathematical restatement looks
natural and general enough:

\begin{center}
\medskip
\framebox{
\parbox[t]{0.8\hsize}{%
%%%%%
\it
A counterfactually definite theory is described by a function (or functions) that map(s) tests onto numbers.The variables
of the function(s) argument(s) must be chosen in a one to one correspondence to physical entities that describe the test(s) and
must be independent variables in the sense that they can be arbitrarily chosen from their respective domains.
%%%%%
}}
\medskip
\end{center}

This definition means that the outcomes of measurements must be described by functions of a set of independent variables. The
definition applies, of course, to the major theories of classical physics, including Einstein's special relativity.
Counterfactual definiteness appears, therefore, as a reasonable and even necessary requirement of classical theories. However,
most importantly, counterfactual definiteness restricts the use of variables to those that can be independently picked from
their respective domains. However, a magnet- or polarizer-orientation, mathematically represented by the variable $\bf j$, cannot be picked independently of the measurement times, which are mathematically represented by $t_n$ and registered by the clocks of the measurement stations. Once a setting is picked at a certain space-time coordinate, no other setting can be linked to that coordinate, because of the relativistic limitations for the movement of massive bodies and the fact that Bell's theory is confined to the realm of Einsteinian physics and, therefore, excludes quantum superpositions. Thus any measurement is related to spatio-temporal equipment changes and the mathematical variables that describe the measurement need to represent the possible physical situations.

Enter probability theory and we certainly cannot use the setting $\bf j$ as a random variable and the measurement time $t$ as
another {\it independent} random variable on the same probability space. The reason for this fact
is rooted in the above explanation and can be further crystallized as follows. It is possible to define the setting $\bf j$ as a random variable on one probability space meaning that we may regard $\bf j$ as a function which assigns to each elementary event $\omega$ of a sample space $\Omega$ a so called realization of $\bf j$ e.g. ${\bf j}(\omega_1) = \mathbf{b}$. It is also possible, at least under very general circumstances, to formulate the measurement times as another random variable $t(\omega')$, where $\omega'$ is an elementary event of a second sample space $\Omega'$. Again, given some specific $\omega_1'$ we obtain a realization e.g. $t(\omega_1') = t_1$.

However, the formation of a product probability space on which both random variables $\bf j$ and $t$ are defined presents now a problem. That space would necessarily contain impossible events (such as different settings for the same
measurement times) with a non-zero product probability measure assigned to them.
These facts can actually be formulated as a theorem stating that setting and
time variables of EPRB experiments cannot be defined on one probability space~\cite{HESS05b}.

Thus, the postulate of counterfactual definiteness in conjunction with the use of a setting variable restricts the independent variables additional to $\bf j$ in the argument of Bell's functions $A$ to a, physically speaking, narrow subset of variables that we denote by $N_B$.
This subset permits the physical description of static properties %,
but cannot handle dynamic properties expressed by space-time dependencies.

As a consequence, the choices that can be made for variables in addition to the setting variable $\bf j$ in Bell's theory are
extremely limited, particularly if these variables are related to space-time (or space and time). This limitation is so severe
that it is impossible to describe general dynamic processes of classical physics with Bell's independent variables. The way to
describe general dynamic processes in Kolmogorov's framework is by using stochastic processes.

To describe a dynamics of EPRB experiments one needs to use two dimensional vector stochastic processes, which
involves several subtleties that, if neglected, lead to incorrect conclusions. A general vector stochastic process is in essence
a vector of random variables, such as $(A_1(t_n), (A_2(t_n), A_3(t_n),...)$, whose statistical properties change in time (we use
here discrete time only). A precise mathematical definition can be found in
Ref.~\cite{BREU02}, pp 11--15.
% [H. P. Breuer and F. Petruccione, "The theory of open quantum systems", Oxford University Press (2002), pp 11-15].
In relation to EPRB experiments we thus consider vectors such
as $(A_1(t_n), A_2(t_n))$.

A first difficulty that is usually encountered is related to the physics of spin measurements.
According to Bohr, the outcomes of measurements on each separate side of the EPRB experiment
are spin-up or spin down with equal likelihood, which appears to suggest stationarity or
time-independence of the random variables $A_1(t_n)$ and $A_2(t_n)$.
Bohr's postulate, however, does not necessitate a time-independence of the statistical correlations between the random
variables. This fact has been explained on the basis of a mathematical model involving time in
Ref.~\cite{HESS15}(pp 55--60)
and demonstrated by actual EPRB related computer experiments~\cite{RAED16a}.

A second difficulty arises from the fact, explained in detail above, that the time and setting related variables of EPRB experiments cannot be treated as independent. This difficulty can be resolved by use of the following two-dimensional system of functions (vector stochastic process) on a probability space $\Omega$:
\begin{equation}
(A_{{\bf j}_n}^{t_n}(\omega), A_{{\bf j'}_n}^{t^\prime_n}(\omega)).
\label{6aug15n1}
\end{equation}
Settings and times are now
included as indexes that are not independent. ${\bf j}_n = {\bf a, b}$ represents the randomly chosen settings at one
measurement place and ${\bf j'}_n = {\bf b, c}$ at the second. $t_n$ as well as $t^\prime_n$ are the respective measurement times.
$n = 1, 2, 3...$ indicates just the number of the experiment. Only one setting can occur at one given time in order to avoid
physical contradictions and incorrect assignments of probability measures. (Note that a generalization of the time-indexes to
space-time ${\bf st }_n $ is straightforward.)

Bell's inequality then transforms to:
\begin{equation}
%\langle
A_{\bf a}^{t_n}(\omega)A_{\bf b}^{t^\prime_n}(\omega)
%\rangle
+
%\langle
A_{\bf a}^{t_k}(\omega)A_{\bf c}^{t^\prime_k}(\omega)
%\rangle
-
%\langle
A_{\bf b}^{t_m}(\omega)A_{\bf c}^{t^\prime_m}(\omega)
%\rangle
\leq 3
,
\label{6aug15n2}
\end{equation}
where the labels $n, k, m$ are the appropriate, all different, experiment
numbers for which the particular settings have been chosen.
Eq~(\ref{6aug15n2})  puts no restrictions on the correlations of
EPRB experiments, because the actual experiments may now be represented by a countable infinite number of different functions instead of the three or four functions used by Bell.

There do exist theorems that appear to prove the validity of Bell's inequality for stochastic processes (the Martingales
discussed in \cite{GILL03a} are just special forms of stochastic processes). These theorems, however, do not use two-dimensional
vector stochastic processes as used in Eq~(\ref{6aug15n1}). They use, instead, counterfactual definiteness in conjunction with
setting variables to arrive at three-, four- or higher dimensional stochastic processes (Martingales). Thus these theorems
cannot encompass dynamic measurement processes~\cite{NIEU13}
and time- (space-time-) related variables, because they would then imply the existence of
events with more than one setting at a given measurement time and, therefore, involve impossible events with non-zero
probability measure. Such theorems apply, therefore, only to the set of variables $N_B$ as defined above and do not apply to
EPRB types of experiments that may involve dynamical processes in the measurement equipment.

It is, therefore, imperative to view EPRB experiments in a different light. A violation of Bell-type inequalities need not be
seen as crossing the border between the reasoning of classical Einstein type of physics and quantum mechanics, but indicating a
possible dynamics in the interactions of particles and measurement equipment. This possible dynamics is what needs to be
investigated, particularly as contrasted to the characterization of the measurement equipment by a completely static symbol~\cite{NIEU11}.

\section{Einstein locality and Bell's reasoning revisited}

Experimentalists have up to now not used Bell's theorem and its implications to search for a many body dynamics of local
equipment, but instead to ``uncover" the instantaneous dynamic influence of remote measurements, the so called quantum non-localities. Some
consider these non-localities to be the most profound development of modern physics \cite{HESS15}. They maintain that the
measurement of the entangled partner causes instantaneous influences over arbitrary distances.

 This search for influences due to distant events is based on the conviction, dating back to Bell's original paper, that
Einstein locality is necessary to derive his inequality. However, this is not the case. Bell's assumption that $\lambda$ is
independent of the setting variable $\bf j$ is already contained in the postulate of counterfactual definiteness and Einstein
locality is not only redundant because of this fact, but does not require at all that $\lambda$ be independent of all settings.
Variables dependent on the local setting and describing local many body interactions with the incoming particles are entirely
permitted and necessary. It is counterfactual definiteness that requires that all additional variables such as $\lambda$ are
independent of the setting variable. But why does our classical theory need to involve the setting variable in the way Bell
has included it? One can use the setting variable as an index together with another index related to or representing space-time. These
indexes are, of course not independent as was pointed out above for stochastic processes.

From these facts we can deduce that Einstein locality is not a necessary condition for Bell's derivation, rather the opposite.
Its correct implementation prevents the derivation of Bell to go forward, as shown in Eq~(\ref{6aug15n2}).

\section{Conclusions}

The major premise for the derivation of Bell's inequality is counterfactual definiteness, which in connection with Bell's use of
setting variables restricts the domain of the variables in the argument of Bell's functions $A$ to a subset $N_B$ of general
physical independent variables. $N_B$ does not include the variables necessary to describe a general dynamics describing many
body interactions in the measurement equipment. Using only the independent variables defined by $N_B$ , it is impossible to find a violation
of Bell's inequality, which therefore represents a demarcation between possible and impossible experience~\cite{BO1862}, not between classical
and quantum physics. For a wider parameter space that permits the description of dynamic processes and includes
space-time coordinates, the validity of Bell-type inequalities cannot be and have not been derived.
This situation is reminiscent
of that with the last theorem of Fermat before 1984. There existed only rather trivial proofs of Fermat's theorem for subsets of
conditions (e.g.  $n=3, 5$), while a general proof was not known until Andrew Wiles supplied it in 1984.
Such more complicated and general proofs of Bell's theorem have not been presented and,
in the authors opinion, are not likely to be presented in the future,
because they would need to remove the use of the setting variable $\bf j$.

\begin{acknowledgments}
We would like to thank the referees for very valuable suggestions that improved the manuscript.
\end{acknowledgments}

%\bibliographystyle{pnas}
%\bibliography{c:/d/papers/all13,c:/d/papers/new15}   % name your BibTeX data base

\end{article}
\end{document}